\documentstyle[prl,aps,preprint]{revtex}

\begin{document}
\draft
\title{Wigner Crystal in One Dimension}

\author{H.J. Schulz}
\address{
Laboratoire de Physique des Solides
\cite{assoc},
 Universit\'{e} Paris-Sud,
91405 Orsay,
France }
\maketitle
\begin{abstract}
A one--dimensional gas of electrons interacting with
long--range Coulomb forces ($V(r) \approx 1/r$) is investigated.
The excitation spectrum consists of separate collective charge and spin
modes, with the charge excitation energies in agreement with RPA
calculations. For arbitrarily weak Coulomb repulsion density
correlations at wavevector $4k_F$ decay extremely
slowly and are best described as those of a one--dimensional Wigner crystal.
Pinning of the Wigner crystal then leads to the nonlinear transport
properties characteristic of CDW. The results allow a consistent
interpretation of the plasmon and spin excitations
observed in one--dimensional semiconductor structures,
and suggest an interpretation of some of the observed features in terms
of ``spinons''.
A possible explanation for nonlinear transport phenomena is given.
\end{abstract}

\narrowtext

The properties of models of one--dimensional interacting electrons
have been studied in great detail. Examples are the so--called ``g--ology''
model of fermions moving in a continuum
\cite{emery_revue_1d,solyom_revue_1d}, or the one--dimensional
Hubbard model \cite{lieb_hubbard_exact,emery_revue_1d,schulz_hubbard_exact}.
In these models, one usually assumes short--ranged (effective)
electron--electron interactions. The so--called ``Luttinger liquid''
\cite{haldane_bosonisation} behaviour in this type
of models is characterized by separation between spin and charge degrees of
freedom and by power--law correlation functions, with interaction--dependent
exponents. Short--range interactions are a reasonable
assumption for applications e.g. to quasi--one--dimensional conductors, where
screeening between adjacent chains leads to effectively short range
interactions within one chain \cite{schulz_coulomb}. However, the situation
can be quite different
if an isolated system of electrons moving in one dimension is considered.
There then is no interchain screening, and the true long--range character of
the Coulomb forces ($V(r) = e^2/r$) needs to be taken into account. This
appears to be
the case, e.g., in certain one--dimensional semiconductor structures, where
the effects of one--dimensional Coulomb forces have indeed been observed
\cite{goni_gaas1d}.

The purpose of the present paper is to investigate the effects of the
long--range Coulomb interaction in a one--dimensional model, using the
bosonization method \cite{emery_revue_1d,solyom_revue_1d}.
This allows in particular a
rather straightforward and asymptotically (for low energies and wavenumbers)
exact description of excitation spectra and correlation functions.
The main conclusion is that the long--range force, even if it is
very weak, leads to a state characterized by quasi--long--range order much
closer to a one--dimensional Wigner crystal \cite{wigner} than to an
electron liquid. The calculations presented here provide a rather simple
microscopic description of
the Wigner crystal, a problem that in higher dimesnions has been difficult
to treat by many--body techniques.

I will start by considering the particular case of one--dimensional
electrons with a linear energy--momentum relation interacting with
long--range Coulomb forces, described by the Hamiltonian
\begin{equation}   \label{ham1}
H= \sum_{k,s} v_F [(k-k_F) a^\dagger_{k,s}a_{k,s} +
(-k-k_F)b^\dagger_{k,s} b_{k,s}] + \frac{1}{2L} \sum_{q} V(q)
\rho_{q} \rho_{-q}
+ H_{bs} \;\;.
\end{equation}
Here $a^\dagger_{k,s}$ ($b^\dagger_{k,s}$) creates a right-- (left--)
moving electron with momentum $k$ and spin projection $s$, $v_F$ is the
Fermi velocity. In the interaction term $\rho_{q}=
\rho_{a,q}+\rho_{b,q}$ is the Fourier component of the total particle
density, and $V(q)$ is the Fourier transform of the interaction
potential. In strictly one dimension, a $1/r$ Coulomb interaction
does not have a Fourier transform because of the divergence for
$r\rightarrow 0$,
however, in a system of finite transverse dimension $d$, the singularity
is cut off a $r \approx d$ \cite{gold_1dplasmon}. Using the approximate form
$V(r) = e^2/\sqrt{r^2+d^2}$, one has $V(q) = 2e^2 K_0(qd)$. Finally, the
backward
scattering term $H_{bs}$ describes processes where particle go from the
right-- to the left--moving branch and vice versa. This involves  a
nonsingular interaction
matrix element at $q \approx 2k_F$, called $g_1$.

The linear energy--momentum relation makes the  model (\ref{ham1})
exactly solvable using standard bosonization methods. Moreover, at least
for weak Coulomb interactions, when states near the Fermi energy play
the major role, linearizing the spectrum is not expected to change the
physics drastically, and one therefore expects that the model
(\ref{ham1}) correctly represents the low--energy physics even for more
realistic bandstructures. The model can be easily solved introducing the
phase fields
\begin{equation}
\phi_\nu(x)  =  -\frac{i\pi}{L}\sum_{p \ne 0} \frac1{p}
e^{-ipx} [\nu_+(p) + \nu_-(p)]
\;\;,
\end{equation}
where $\nu=\rho,\sigma$, and $\rho_r(p)$ ($\sigma_r(p)$) are the usual
charge (spin) density operators for right--($r=+$) and left--($r=-$)
going fermions. The Hamiltonian then decomposes into commuting parts for the
charge and spin degrees of freedom. The charge part takes the simple
quadratic form
\begin{eqnarray}     \nonumber
 H_\rho & = & \frac{v_F}{2 \pi} \int dx \left[ \pi^2 (1 + \tilde{g}_1)
\Pi_\rho^2
       + (1-\tilde{g}_1) (\partial_x \phi_\rho )^2 \right] \\
\label{ham2}
& + & \frac{1}{\pi^2} \int dx dx'V(x-x') \partial_x \phi_\rho
\partial_{x'} \phi_\rho
\end{eqnarray}
Here $\Pi_\rho$ is the momentum density conjugate to $\phi_\rho$, and
$\tilde{g}_1 = g_1/(2\pi v_F)$. The Hamiltonian (\ref{ham2}) is quadratic
in the bosonic fields and therefore can be diagonalized
straightforwardly. The
elementary excitations then are found to be charge oscillations (plasmons),
with energy--momentum relation
\begin{equation}            \label{pla}
\omega_\rho (q) = v_F |q| [(1 + \tilde{g}_1)(1 -
\tilde{g}_1+2\tilde{V}(q))]^{1/2} \end{equation}
where $ \tilde{V}(q) = V(q) / (\pi v_F)$. The long--wavelength form,
$\omega_\rho (q) \approx |q^2 \ln q|^{1/2}$, agrees with RPA calculations
\cite{gold_1dplasmon,li_1dplasmon}, however, the effect of $g_1$, which is
a short--range exchange contribution, are usually neglected in those
calculations.

The spin part of the Hamiltonian does not involve the long--range part of
the interaction and only depends on the backward scattering amplitude $g_1$.
For repulsive interaction, the long--wavelength spin excitations then are
described by a Hamiltonian similar to the first term in (\ref{ham2}), giving
rise to collective spin oscillations with $\omega_\sigma(q) = u_\sigma |q|$,
and spin wave velocity $u_\sigma = v_F \sqrt{1-\tilde{g}_1^2}$. Together
with the charge oscillations (\ref{pla}), these excitations are the complete
spectrum of the model.

The bosonization method makes the calculation of correlation functions
rather straightforward. Here, the charge--charge correlations are of
particular interest. Using the expression
\begin{eqnarray} \nonumber
\rho (x) &= &- (\sqrt2/\pi)\partial_x \phi_\rho (x) \\
 &+& \frac{1}{2\pi
\alpha} e^{2ik_Fx} e^{-i\sqrt2 \phi_\rho(x)} \cos[\sqrt2 \phi_\sigma(x)]
+ cst.  e^{4ik_Fx} e^{-i\sqrt8 \phi_\rho(x)} + h.c.
\end{eqnarray}
the evaluation of the charge correlation function reduces to the calculation
of averages of the type
\begin{equation}    \label{av}
\langle (\phi_\rho(x) - \phi_\rho(0))^2 \rangle = \int_0^\infty
\frac{dq}{q}
\left[\frac{1 + \tilde{g}_1}{1-\tilde{g}_1+2\tilde{V}(q)}\right]^{1/2}
(1 - \cos qx) \approx c_2 \sqrt{\ln x} \;\;,
\end{equation}
with $c_2 = \sqrt{(1+\tilde{g}_1) \pi v_F/e^2}$.
One thus obtains
\begin{equation}     \label{corr}
\langle \rho(x) \rho(0) \rangle = A_1 \cos(2k_Fx)
\exp(- c_2 \sqrt{\ln x})
/x + A_2 \cos(4k_Fx) \exp(- 4 c_2 \sqrt{\ln x}) + ... \;\;,
\end{equation}
where $A_{1,2}$ are interaction dependent constants, and only the most
slowly decaying Fourier components are exhibited.
The most interesting point here is the extremely slow decay (much slower
than any power law!) of the $4k_F$ component, showing an incipient charge
density wave at wavevector $4k_F$ (instead of the usual $2k_F$ of the
Peierls instability). This slow decay should be compared with the case of
short--range interactions, where the $2k_F$ and $4k_F$ components deacay as
with the power laws $x^{-1-K_\rho}$ and $x^{-4K_\rho}$, respectively, with
an interaction--dependent constant $K_\rho$
\cite{emery_revue_1d,solyom_revue_1d,schulz_hubbard_exact}.
The $4k_F$ oscillation period
is exactly the average interparticle spacing, i.e. the structure is that
expected for a one--dimensional {\em Wigner crystal}. Of course, because of
the one--dimensional nature of the model, there is no true long--range
order, however, the extremely slow decay of the $4k_F$ oscillation would
produce strong quasi--Bragg peaks in a scattering experiment. It is
worthwhile to point out that this $4k_F$ contribution arises even if the
Coulomb interaction is extremely weak and depends only on the long--range
character of the interaction. On the other
hand, any $2k_F$ scattering is considerably weaker, due to the $1/x$
prefactor in (\ref{corr}) which has its origin in the contribution of spin
fluctuations.

Other correlation functions are easily obtained. For example, the spin--spin
correlations are
\begin{equation}
\langle \mbox{\boldmath ${S}$}(x) \cdot \mbox{\boldmath ${S}$}(0) \rangle
\approx B_1
\cos(2k_Fx) \exp(- c_2 \sqrt{\ln x})/x + ...
\end{equation}
where there is no $4k_F$ component. On the other hand, correlation functions
that involve operators changing the total number of particles (e.g. the
single particle Green's function) decay like $\exp[- cst.(\ln x)^{-3/2}]$,
i.e. {\em faster} than any power law. This in particular means that the
momentum distribution function $n_k$ and all its derivatives are continuous
at $k_F$, and there is only an essential singularity at $k_F$. The
calculations are also straightforwardly generalized to finite frequency and
temperature \cite{emery_revue_1d,solyom_revue_1d}, however the rather
complicated formulae are not of immediate interest here.

The presence of metallic screening changes the above behaviour: a finite
screening length $\xi_s$ would lead to a saturation of $V(q)$ for $q
\rightarrow 0$ at $2e^2 \ln (\xi_s/d)$. One than would have, for $x >
\xi_s$,
power--law decay of the type discussed above for short--range interactions,
with $K_\rho \approx 1/\sqrt{\ln \xi_s}$. On the other hand, if the
interaction potential decays more slowly than $1/r$ (a rather hypothetical
case), the integral (\ref{av}) remains finite for $x \rightarrow \infty$,
and therefore there then is real long--range order of the Wigner crystal
type.

It is instructive to compare the above result (\ref{corr}), obtained in the
limit of weak Coulomb interactions, with the case of strong repulsion (or,
equivalently, heavy particles). The configuration of minimum potential
energy is one of a chain of equidistant particles with lattice constant $a$,
and quantum effects are expected to lead only to small oscillations in the
distances between particles. The Hamiltoniam then is
\begin{equation}
H = \sum_l \frac{p_l^2}{2m} + \frac14 \sum_{l \neq m} V''(ma)
(u_l - u_{l+m})^2 \;\;,
\end{equation}
where $u_l$ is the deviation of particle $l$ from its equilibrium position.
In the long--wavelength limit, the oscillation of this lattice have energy
$\omega(q) = \sqrt{2/(ma)} e q |\ln (qa)|^{1/2}$. The most slowly
decaying part of the density--density correlation function then is
\begin{equation}   \label{corr2}
    \langle \rho(x) \rho(0) \rangle \approx
\cos(2 \pi x/a) \exp \left[ - \frac{4\pi}{(2 m e^2 a)^{1/2}} \sqrt{\ln
x}\right] \;\;.
\end{equation}
Noticing that $k_F = \pi/(2 a)$, one observes that the results (\ref{corr})
and (\ref{corr2}) are (for $g_1=0$) identical as far as the
long--distance asymptotics are concerned,
{\em including the constants in the exponentials}.
Eq. (\ref{corr}) was obtained in the weak interaction
limit, whereas (\ref{corr2}) applies for strong Coulomb forces.
Similarly, the small--$q$ limit of the charge excitation energies is
identical. We
thus are lead to the rather remarkable conclusion that the
long--distance behaviour of
correlation functions is independent of the strength of the Coulomb
repulsion, provided the interaction is truly long--ranged.

In recent experiments, one--dimensional structures with two partially
filled
subbands have been investigated \cite{goni_gaas1d}. If only the long--range
part of the Coulomb interaction is considered, the appropriate
generalization of the model (\ref{ham2}) to that case is described by the
Hamiltonian \begin{eqnarray}     \nonumber
 H_\rho & = & \sum_{i=0,1} \frac{v_i}{2 \pi} \int dx \left[ \pi^2
 \Pi_{\rho,i}^2
       +  (\partial_x \phi_{\rho,i} )^2 \right] \\
%\label{ham3}
\nonumber
&& +  \frac{1}{\pi^2} \int dx dx'V(x-x') \partial_x (\phi_{\rho,0}+
\phi_{\rho,1})
\partial_{x'} (\phi_{\rho,0}+ \phi_{\rho,1}) \;\;,
\end{eqnarray}
where $v_i$ is the Fermi velocity of band $i$, and $\phi_{\rho,i},
\Pi_{\rho,i}$ are the charge fields of band $i$. In the
long--wavelength limit the charge oscillation eigenmodes have energies
\begin{displaymath}
\omega_+(q) =  |q| \sqrt{2(v_0+v_1)V(q)/\pi} \quad , \quad
\quad
\omega_-(q) = \sqrt{v_0 v_1} |q|  \;\;.
\end{displaymath}
The $\omega_+$ mode represents in phase oscillations of the two bands
and has the typical $|q^2 \ln q|^{1/2}$ behaviour of one--dimensional
plasmons, whereas the $\omega_-$ mode is an out--of--phase oscillation.
In addition there are two spin modes, at energies $v_{0,1} |q|$. If the
various possible interaction processes involving momentum tranfer of
order $2k_{F,i}$ are included, the velocities of these modes are
renormalized, similar to the effect of $g_1$ in the one--band model
above.

For the density correlations of the two--band model I find
\begin{eqnarray*}
\langle \rho(x) \rho(0) \rangle & = &
C_2 \cos[4(k_{F,0}+ k_{F,1})x] \exp (-4 c_2 \sqrt{\ln x})  \\
&& + x^{-K}\sum_{i=0,1} C_i \cos(4k_{F,i}x) \exp(-c_i\sqrt{\ln x})
\end{eqnarray*}
where now $c_2 = \sqrt{\pi(v_0+v_1)/e^2}$, $c_i = 4v_i^2
c_2/(v_0+v_1)^2$, $K=4\sqrt{v_0v_1}/(v_0+v_1)$, and again only the
most slowly decaying Fourier components are exhibited. The most slowly
decaying part of (\ref{corr2}) ($q=4(k_{F,0}+ k_{F,1})$) is again the
one that corresponds to Wigner--crystal type ordering, e.g. the
electrons
order approximately equidistantly. In fact, this type of ordering
is determined only by the $\omega_+$ mode, whereas all other Fourier
componets
contain contributions from the $\omega_-$ mode, which lead to power law
decay.

To compare the present results with experiment \cite{goni_gaas1d} one can
first notice that, provided that $2k_F d < 1$ and including the background
dielectric screening, one has $\tilde{g}_1 < 0.2$, and consequently
to within a few percent $u_\sigma = v_F$, i.e. the triplet spin mode
(``SDE'') is expected at $v_F q$, as experimentally observed. Further,
in the experimentally accessible
range $q < 0.2k_F$ the plasmon energies found here are indistinguishable
from RPA results, and thus the present results provide a good fit to the
experimental plasmon dispersion.

More difficult to explain is the
extra feature which has been interpreted as an electron--hole
continuum (``SPE'')
\cite{goni_gaas1d}: in fact in the present model with its linear electron
dispersion relation, there is no such continuum (and it would not exist in
an RPA calculation either). However, the model offers an alternate
possibility:
 together with the triplet spin mode, there
is also a {\em singlet} mode \cite{rem1}.
The existence of the singlet mode is a consequence of
spin--charge separation in one--dimensional
fermion systems, and in particular it is degenerate with the triplet mode.
This mode can be found e.g. in energy density correlation
functions (as opposed to the spin mode, which appears as a pole of the spin
density correlation function), and therefore also is expected to be seen in
the polarized Raman spectra. This interpretation requires the SDE
and SPE features to appear at  the same energy, which seems to be
consistent with the results published in ref.\cite{goni_gaas1d}.
It is noteworhty that, if correct, this interpretation would mean that
these results constitute the first direct
spectroscopic evidence for the existence of individual spin--1/2 objects
(``spinons''): the existence of degenerate triplet and singlet mode
implies that they are both build up from non--interacting spin--1/2
excitations.

One might argue that the existence of the particle--hole continuum is
due to effects of band curvature, which is neglected in the present
model. RPA calculations including band curvature certainly predict
both a plasmon and a particle--hole continuum
\cite{gold_1dplasmon,li_1dplasmon}.
However, within RPA the total spectral weight for
the continuum is about two orders of magnitude smaller than that of the
plasmon, whereas in ref.\cite{goni_gaas1d} plasmon and SPE have
comparable weight. Moreover, in exactly solved one--dimensional models
like the Hubbard model \cite
{lieb_hubbard_exact}
, one finds both a plasmon--like collective
mode
and the singlet mode discussed above, but no separate particle--hole
continuum \cite{rem2}.
There thus seems to be little theoretical evidence in favor of an
interpretation of the SPE feature in terms of
a particle--hole continuum.

The nearly long--range Wigner crystal type order should have important
consequences for transport properties: in fact, in the presence of disorder,
a classical charge density wave (which has real long--range order at $T=0$)
becomes disordered \cite{lee_fuku}, with a ``pinning length''
describing the decay of spatial correlations given by
\begin{equation}          \label{pin}
\xi_{pin} \approx ((V/v_F)^2n)^{-1/3} \;\;,
\end{equation}
where $n$ is the density of impurities, and $V$ the Fourier
component of the impurity potential at the wavevector of the CDW ($4k_F$ in
our case). Inclusion of quantum effects in systems with short range
interaction only leads to corrections to the exponent $1/3$ in (\ref{pin})
\cite{giamarchi_loc}. Following the same arguments, I expect (\ref{pin})
to be valid for the Coulomb system too (up to logarithmic corrections),
i.e. as far as low--frequency phenomena are concerned, the system of
electrons interacting with Coulomb forces {\em behaves like a classical
charge density wave}. In particular, all the unusual dynamical properties
associated with nonlinear transport in CDW systems should also occur in the
one--dimensional electron system.

At finite temperature, thermal agitation can become sufficiently strong
to depin a CDW. In the present case, this is expected to happen when the
thermal correlation length, $\xi_T$, in the absence of impurities, given
by $\omega_\rho(1/\xi_T) \approx T$, becomes shorter than $\xi_{pin}$.

Nonlinear current--voltage relations characteristic of
CDW transport have been observed in one--dimensional semiconductor
structures
\cite{scott-thomas_1d}.
%\cite{scott-thomas_1d,field_1d,meirav_gaas_1d}.
In these experiments,
strong variations of the {\em linear} conductance with carrier density have
been interpreted in terms of the Coulomb blockade
%\cite{vanhouten_coulomb,kastner_coulomb},
\cite{vanhouten_coulomb},
implying the existence of a pair of strong impurity potentials.
It seems certainly conceivable
that the {\em nonlinear} behaviour should be due to collective motion of a
pinned CDW, possibly in parts of the sample where there is only a weak
random potential. The fact
 that the order of magnitude of the nonlinear conductance is
independent of the value of the linear conductance suggests that
different mechanisms are involved in the two phenomena.
The
present calculation then demonstrates that Coulomb interactions do not have
to be particularly strong to create nearly classical CDW type behaviour.
It would clearly be
interesting to investigate the possible interplay between Coulomb
blockade and CDW--like behaviour.

In conclusion, using the bosonization technique a consistent
microscopic picture of
the excitation spectrum and the correlation functions of a one--dimensional
electron gas interacting with long--range Coulomb forces has been obtained.
The density correlations are those of a nearly perfect one--dimensional
Wigner crystal. The results are in agreement with
excitation spectra observed in one--dimensional
semiconductor structures and provide a possible explanation for
nonlinear transport properties.

\noindent
{\em Acknowledgement --} I am grateful to D.S. Fisher for pointing out the
work reported in ref.\cite{scott-thomas_1d}.

%\bibliography{revues,1dtheory,semic,wig1d}

\begin{thebibliography}{10}

\bibitem[*]{assoc} Laboratoire associ\'e au CNRS.

\bibitem{emery_revue_1d}
V.~J. Emery,  in {\em Highly Conducting One-Dimensional Solids}, edited by
  J.~T.~D. et~al. (Plenum, New York, 1979), p.\ 327.

\bibitem{solyom_revue_1d}
J. S\'olyom, Adv. Phys. {\bf 28},  209  (1979).

\bibitem{lieb_hubbard_exact}
E.~H. Lieb and F.~Y. Wu, Phys. Rev. Lett. {\bf 20},  1445  (1968).

\bibitem{schulz_hubbard_exact}
H.~J. Schulz, Phys. Rev. Lett. {\bf 64},  2831  (1990).

\bibitem{haldane_bosonisation}
F.~D.~M. Haldane, J. Phys. C {\bf 14},  2585  (1981).

\bibitem{schulz_coulomb}
H.~J. Schulz, J. Phys. C {\bf 16},  6769  (1983).

\bibitem{goni_gaas1d}
A.~R. Goni {\it et~al.}, Phys. Rev. Lett. {\bf 67},  3298  (1991).

\bibitem{wigner}
E. Wigner, Phys. Rev. {\bf 46},  1002  (1934).

\bibitem{gold_1dplasmon}
A. Gold and A. Ghazali, Phys. Rev. B {\bf 41},  7626  (1990).

\bibitem{li_1dplasmon}
Q.~P. Li and S.~D. Sarma, Phys. Rev. B {\bf 43},  11768  (1991).

\bibitem{rem1}
I am grateful to C. Stafford for interesting comments on this point.

\bibitem{rem2}
H.J. Schulz, unpublished.

\bibitem{lee_fuku}
P.~A. Lee and H. Fukuyama, Phys. Rev. B {\bf 17},  535  (1978).

\bibitem{giamarchi_loc}
T. Giamarchi and H.~J. Schulz, Phys. Rev. B {\bf 37},  325  (1988).

\bibitem{scott-thomas_1d}
J.~H.~F. Scott-Thomas {\it et~al.}, Phys. Rev. Lett. {\bf 62},  583  (1989);
S.~B. Field {\it et~al.}, Phys. Rev. B {\bf 42},  3523  (1990);
U. Meirav, M.~A. Kastner, M. Heilblum, and S.~J. Wind, Phys. Rev. B {\bf 40},
  5871  (1989).

\bibitem{vanhouten_coulomb}
H. van Houten and C.~W.~J. Beenakker, Phys. Rev. Lett. {\bf 63},  1893
(1989); M.~A. Kastner, Rev. Mod. Phys. {\bf 64},  849  (1992).

\end{thebibliography}
%\bibliographystyle{prsty}

\end{document}